\documentclass[prd,preprintnumbers,floatfix,aps,notitlepage,nofootinbib,twocolumn,showpacs,amssymb]{revtex4-2}
\usepackage{amsmath,amsfonts,bm}
\usepackage{graphicx}
\usepackage{cancel}
\usepackage[colorlinks, linkcolor={red},citecolor={blue}]{hyperref}
\usepackage{xcolor}

\begin{document}
\title{Interior evolution of Regular Schwarzschild Black Holes}

\author{J Ovalle}
\email[]{jorge.ovalle@physics.slu.cz}
\affiliation{Research Centre for Theoretical Physics and Astrophysics,
	Institute of Physics, Silesian University in Opava, CZ-746 01 Opava,
	Czech Republic}	

\affiliation{Universidad de Tarapac\'a, Avenida Luis Emilio Recabarren 2477, Iquique, Chile
}

\begin{abstract}
\noindent We present an exact, purely geometric {description of the interior evolution} of Schwarzschild black holes, formulated without invoking any specific gravitational theory and free of additional charges beyond the total mass ${\cal M}$. We show that {the time dependent interior}  generically produces new singularities, absent in the static case, whose resolution imposes highly restrictive conditions on gravitational collapse.

		\end{abstract} 
\maketitle
%
%
%
\section{Introduction}
\noindent Among the various strategies for circumventing the celebrated Penrose singularity theorem~\cite{Penrose:1964wq,Hawking:1970zqf,Hawking:1973uf}, the most commonly explored involves allowing the existence of an inner horizon: one typically constructs a regular black hole (BH) solution~\cite{Bardeen:1968qtr,Dymnikova:1992ux,Hayward:2005gi,BenAchour:2020gon,Bonanno:2023rzk}, only to find that it necessarily contains at least one inner horizon, or Cauchy horizon, thereby breaking global hyperbolicity. As expected, given the theorem's generality, it does not specify the {evolution} of gravitational collapse, something essential if we aim to develop a physically relevant regularization mechanism that is, as far as possible, independent of spacetime symmetry. {Modeling collapse and singularity formation in a simple analytical form is, however, notoriously difficult, even under maximal symmetry, except in highly idealized cases~\cite{Joshi:2008zz}. Nevertheless, in this work we overcome this obstacle by adopting two well-motivated premises:}

 (i) The event horizon must precede the singularity, as required by weak cosmic censorship~\cite{Penrose:1969pc}. For a finite stage before full collapse, the infalling matter settles into a regular configuration that necessarily develops an inner horizon; in static or stationary cases this inner horizon coincides with a Cauchy horizon, for which extensive evidence shows generic instability~\cite{Penrose:1968,Poisson:1989zz,Ori:1991zz,Carballo-Rubio:2018pmi,Carballo-Rubio:2021bpr}. 
 
 (ii) Before total collapse occurs, a stage marked by maximal geometric simplicity, we may reasonably expect the internal geometry of the BH to be highly complex. This complexity, however, will be assumed to remain consistent with the absence of primary hair: a single parameter ${\cal M}$ in the static case, and two parameters $\{{\cal M},\,a\}$ in the stationary case, representing the total mass and angular momentum of the configuration, respectively. In short, any internal structure should be encoded entirely in secondary hair. This requirement implies one of two possibilities: (a) the system remains within the domain of general relativity, even under extreme curvature, or (b) if a new theory is invoked, its formulation must remain strictly geometric and refrain from introducing additional charges.

Building on the two premises outlined above, this work presents a simple analytical description of the final stage of spherically symmetric gravitational collapse, namely the formation of a nonsingular Schwarzschild BH, developed within a purely geometric framework, without assuming any specific gravitational theory. We first identify an infinite family of previously overlooked regular solutions, revealing the remarkable structural richness of the Schwarzschild interior. By introducing {time dependence} into these solutions, we uncover potential singularities that arise only in the {evolving} case, absent in the static scenario, whose resolution shows that collapse cannot proceed arbitrarily but is instead subject to highly restrictive conditions. We further demonstrate, without resorting to perturbative methods~\cite{Poisson:1989zz,Ori:1991zz}, that the region beyond the inner horizon inevitably develops singularities, in full agreement with the strong cosmic censorship conjecture~\cite{Dafermos:2003wr,Penrose:1969pc,Dafermos:2017dbw,Hollands:2019whz}. Finally, by exploiting the simplicity of these configurations, we propose a tractable regularization mechanism that provides both an exact analytical account of regular BH formation and a detailed description of the transition between Schwarzschild's singular and regular configurations.

 \section{Regular Schwarzschild BHs.}
 \noindent
We begin with the interior of the Schwarzschild BH~\cite{Ovalle:2024wtv, Mars:1996khm,Maeda:2024tpl}.
The complete spacetime is described by the line element
\begin{equation}
	\label{metric}
	ds^{2}
	=
	-f(r)\,dt^{2}
	+\frac{dr^2}{f(r)}
	+r^2\,d\Omega^2\ ,
\end{equation}
where
\begin{equation}
	\label{mtransform}
	f(r)=\left\{
	\begin{array}{l}
		1-\frac{2\,m(r)}{r}
		\ ,
		\quad
		{\rm for}\
		0< r \leq h
		\\
		\\
		1-\frac{2\,{\cal M}}{r}
		\ ,
		\quad
		{\rm for}\
		r\geq{h}
		\ .
	\end{array}
	\right.
\end{equation}
The function $m(r)$ in~\eqref{mtransform} stands for the Misner-Sharp mass and
\begin{equation}
	\label{cond1}
	{\cal M}\equiv\,m(r)\rvert_{r=h}=\frac{h}{2}
\end{equation} 
is the Arnowitt-Deser-Misner mass, where $h$ is the event horizon. Notice that the metric function $m(r)$ remains unknown for $r<h$. We can see that the Schwarzschild solution~\cite{Schwarzschild:1916,Eddington:1924pmh,Lemaitre:1933gd,Finkelstein:1958zz,Kruskal:1959vx,Szekeres:1960gm,Dafermos:2021cbw} {is recovered as the particular case $m(r)={\cal M}$ throughout the entire domain $0<\,r\leq\infty\ .$}
To ensure the smooth continuity of the metric~\eqref{metric} across the horizon $r=h$, the mass function $m(r)$ must satisfy
\begin{equation}
	\label{cond2}
	m(h)={\cal M}\ ;\,\,\,\,\,\,\,m'(h)=0\ ,
\end{equation}
where $F(h)\equiv\,F(r)\big\rvert_{r=h}$ for {any function} $F(r)$. 
 
 Now, consider the mass function of the interior metric in~\eqref{mtransform}, given by
 \begin{equation}
 	\label{M}
 	m
 	=
 	M
 	-
 	\frac{Q^2}{2\,r}
 	+
 	\frac{1}{2}\,\sum_{n=2}^{\infty}\,\frac{C_n\,r^{n+1}}{(n+1)(n+2)}
 	\ ,
 	\quad
 	n\in\mathbb{N}
 	\ ,
 \end{equation}
 where $M$ and $Q$ are integration constants that may eventually be identified with the Schwarzschild mass and the Reissner-Nordstr\"{o}m charge, respectively~\cite{Ovalle:2024wtv}. Nonetheless, in order to guarantee regularity, {we impose $M=Q=0$, in accordance with premise (i) stated in the Introduction.} While the general series solution in~\eqref{M} is highly effective~\cite{Casadio:2024fol,Aoki:2024dyr,Casadio:2025pun}, it still depends implicitly on $h=2\,{\cal M}$ through the coefficients $C_n$. It would be far more useful to derive a general solution that expresses this dependence explicitly. Such a formulation would allow for a clearer and more direct analysis of the Schwarzschild BH interior. To accomplish this, we consider a generic solution expressed as a superposition of $N$ distinct configurations~\cite{Ovalle:2017fgl,Ovalle:2019qyi}, embedded in a de Sitter background with cosmological constant $\sim\,C_3$,
 \begin{equation}
 	\label{mpoly}
 	m(r)
 	=
 	C_3\,r^3
 	+
 	\underbrace{C_l\,r^l
 		+C_n\,r^n
 		+C_p\,r^p+...}_{\text{$N$ terms}}
 	\ ,
 \end{equation}
 {where the coefficients $C_s$ depend only on the total mass ${\cal M}$, in accordance with premise (ii) stated in the Introduction. They are determined by imposing condition~\eqref{cond1} together with}
 \begin{equation}
 	\label{cond-n}
 	\frac{d^n m}{dr^n}(h)
 	=
 	0
 	\ ,
 \end{equation}
 for all $1\leq n\leq N$. From~\eqref{cond-n} it follows that $m(r)$ is $N$-times differentiable at the event horizon $r=h$. Consequently, both $m(r)$ in~\eqref{mtransform} and the metric~\eqref{metric} belong to the class ${\cal C}^N$ for $r>0$. Applying this method, the interior Schwarzschild mass function in~\eqref{mtransform} takes the form
 \begin{eqnarray}
 	\label{minfi}
 	&&	m(r)=\frac{r}{2}\left[\left(\frac{r}{h}\right)^2\,\prod_{i=1}^{N}\frac{n_i+1}{n_i-2}\right.\nonumber\\
 	&&\left.+3(-1)^N\,\sum_{k=1}^{N}\frac{1}{n_k-2}\left(\frac{r}{h}\right)^{n_k}\prod_{\substack{i=1\\i\neq k}}^{N}\frac{n_i+1}{n_k-n_i}\right]\ ,
 \end{eqnarray} 
which yields the corresponding Schwarzschild interior metric function
  \begin{eqnarray}
 	\label{finfi}
 	&&	f(r)=1-\left[\left(\frac{r}{h}\right)^2\,\prod_{i=1}^{N}\frac{n_i+1}{n_i-2}\right.\nonumber\\
 	&&\left.+3(-1)^N\,\sum_{k=1}^{N}\frac{1}{n_k-2}\left(\frac{r}{h}\right)^{n_k}\prod_{\substack{i=1\\i\neq k}}^{N}\frac{n_i+1}{n_k-n_i}\right]\ ,
 \end{eqnarray} 
  where $2<n_i\in\mathbb{N}$. 
   For each fixed $N$, the solution is specified by $N$ parameters $n_i$ (not hairs), parametrizing an infinite family of regular BH geometries. Extending the domain to $n_i \in [-2,2]$ yields BHs with integrable singularities~\cite{Lukash:2013ts,Ovalle:2023vvu,Arrechea:2025fkk}. In particular, if there exists an index $i$ such that $n_i=-1$, one recovers the standard Schwarzschild solution. 
   
    {We stress that the mass function $m(r)$ in Eq.~\eqref{minfi} provides a minimal ${\cal C}^{N}$ construction that simultaneously enforces regularity at the origin, satisfies the condition $m'(r)>0$ required by the null convergence condition, and guarantees smooth matching at $r=h$. In the limit $N\to\infty$, this construction naturally extends to a series representation of smooth mass functions, thereby encompassing a broad and physically meaningful class of gravitational collapse profiles.}
   
   {Regarding the roots of $f(r)=0$ for a fixed $N$, the total number of roots will be equal to the largest exponent $n_i$ in the set $n_i=\{n_1, n_2,\ .\ .\ .n_N\}$. Nevertheless, independently of the value of $N$, there are always only two real roots within the region $0<r\leq\,h$. One corresponds to the event horizon $r=h$, while the other defines the inner, or Cauchy, horizon $h_c<h$.}
   
   As a concrete illustration, let us consider the simplest case with $N=1$, \footnote{In this case the term $i\neq\,k$ produces an empty product and therefore evaluates to $1$.} 
  which takes the explicit form
  \begin{equation}
 	\label{m1}
 	m(r)=\frac{r}{2(n-2)}\left[\frac{r^2}{h^2}\left(n+1\right)-3\left(\frac{r}{h}\right)^n\right]\ ;\,\,\,n>2\ ,
 \end{equation}
 and the case $N=2$
 \begin{eqnarray}
 	\label{Mr}
 	m(r)=&&\frac{r}{2}\left[\frac{(n+1)(l+1)}{(n-2)(l-2)}\left(\frac{r}{h}\right)^2+\frac{3\,(l+1)}{(n-2)(n-l)}\left(\frac{r}{h}\right)^n\right.
 	\nonumber\\
 	&&\left.+\frac{3\,(n+1)}{(l-2)(l-n)}\left(\frac{r}{h}\right)^l \right]\ ;\,\,\,l>n>2\,\in\mathbb{N}\ .
 \end{eqnarray}
We can see that both regular solutions~\eqref{m1} and~\eqref{Mr}, originally derived in Ref.~\cite{Ovalle:2024wtv} and subsequently analyzed in Ref.~\cite{Casadio:2025pun}, represent special cases of the solution~\eqref{finfi}. Regarding the scalar curvature, it takes the form
   \begin{eqnarray}
  	\label{Rinfi}
   \hspace*{-6mm}	&&	R(r)=\frac{3}{h^2}\left[4\,\prod_{i=1}^{N}\frac{n_i+1}{n_i-2}\right.\nonumber\\
   \hspace*{-6mm}	&&\left.+(-1)^N\,\sum_{k=1}^{N}\frac{(n_k+1)(n_k+2)}{(n_k-2)}\left(\frac{r}{h}\right)^{n_k-2}\prod_{\substack{i=1\\i\neq k}}^{N}\frac{n_i+1}{n_k-n_i}\right]
  \end{eqnarray} 
with the property that $R(h)=0$ for $N>1$. We emphasize that these BHs possess a single inner horizon $h_c$. Finally, we highlight two important aspects of the solution~\eqref{finfi}, namely, (i) it depends only on the mass ${\cal M}$ of the configuration (no hairs) and (ii) despite it contains only a single parameter ${\cal M}$, the solution has an (quasi) extremal BH with $h_c\sim\,h$. This occurs when  	$n_k>>1\,\,\forall\,k$, yielding	
	$m(r)\sim\frac{r^3}{2h^2}$. 
	
In connection with the latter property, a direct inspection of the metric function in Eq.~\eqref{finfi}, or equivalently the scalar curvature in Eq.~\eqref{Rinfi}, yields
the explicit expression for the effective cosmological constant,
\begin{equation}
	\label{Lambda-effec}
	\Lambda_{eff}=\frac{3}{h^2}\prod_{i=1}^{N}\frac{n_i+1}{n_i-2}\ .
\end{equation}
\begin{table*}
	{
	\caption{Behavior of the Kretschmann scalar $\mathcal{K}$ for various values of the parameter $n$, corresponding to integrable singularities, in the case $N=1$ of Eq.~\eqref{finfi}. The evolution from any configuration with $n(v)>2$, which is nonsingular, toward $n=-1$, corresponding to the Schwarzschild solution, necessarily passes through a timelike singularity before reaching a spacelike one.
		\label{tab}}
	\begin{ruledtabular}
		\begin{tabular}{ c c c c c}
			$n$ & Scaling &Notes& $\Lambda$ &Singularity
			\\
			\hline\hline
			$2$
			&
			$\mathcal{K} \sim [\log(r/h)]^2/h^4$  & $0<h_{\mathrm{c}}<h$& $\text{AdS}$ &Timelike
			\\
			\hline
			$1$
			&
			$\mathcal{K} \sim (hr)^{-2}$   &$0<h_{\mathrm{c}}<h$ &$\text{AdS}$ &Timelike
			\\
			\hline
			$0$
			&
			$\mathcal{K} \sim r^{-4}$  &Inner horizon $h_{\mathrm{c}}$ vanishes &$\text{AdS}$  &Spacelike
			\\ 
			\hline
			$-1$ & 
			$\mathcal{K} \sim h^2 / r^6$
			&Schwarzschild & $0$ &Spacelike
			\\ 
			\hline
			$-2$ & 
			$\mathcal{K} \sim h^4 / r^8$ &  dS+Conformal solution & $\text{dS}$ &Spacelike
			\\  
		\end{tabular}
	\end{ruledtabular}
}
\end{table*}
 As $n$ grows large $h_{\rm c}\sim\,h$, implying that the configuration approaches an {quasi} extremal Schwarzschild BH with a de Sitter interior, i.e., 
\begin{equation}
	\label{extremal}
	m(r) \to \frac{r^3}{2h^2}, \quad
	f(r) \to 1 - \frac{r^2}{h^2}, \quad h_{\rm c} \to h \quad \text{as} \quad n_i \to \infty\ .
\end{equation}
This de Sitter saturation is particularly significant because, unlike all other regular BH solutions where the de Sitter core remains localized near the origin $r\sim0$, here it undergoes unbounded growth $({n_i\to\infty})$ until occupying the complete interior spacetime $(0\leq\,r\leq\,h)$. 

Finally, notice that the surface gravity $\kappa$ for the metric function~\eqref{finfi}, given by
\begin{equation}
	\label{kappa}
	\kappa = \frac{1}{2} f'(h) = \frac{1}{2h}\ ,
\end{equation}
is completely independent of the parameters $n_i$. This result reveals that, regardless of the interior's complexity, the horizon thermodynamics remains purely Schwarzschil-like, entirely governed by $h$ alone. {The strict limiting case $m(r)=r^3/(2h^2)$ in Eq.~\eqref{extremal} provides the sole exception to this result, exhibiting a de Sitter geometry with surface gravity $\kappa=1/h$. Here, while the spacetime remains continuous at the horizon, it fails to be differentiable, a property that directly explains the discontinuous jump in surface gravity from the interior de Sitter value ($\kappa=1/h$) to the exterior Schwarzschild value ($\kappa=1/2h$). 
{We conclude by emphasizing that the surface gravity associated with the inner horizon
\begin{equation}
	\label{kappa2}
	\kappa_{\rm c} = \frac{1}{2} f'(h_{\rm c})
\end{equation}
depends explicitly on the parameters $n_i$ and remains nonvanishing for the regular configurations considered here. Hence, the fact that the surface gravity of the outer horizon remains Schwarzschild-like, $\kappa=1/2h$, does not suppress the standard Cauchy-horizon instability, which is instead controlled by $\lvert{\kappa_{\rm c}}\rvert$. We therefore expect the exponential blueshift associated with the inner horizon to persist also in the quasi-extremal regime, including configurations approaching the unbounded de Sitter core in~\eqref{extremal}. However, as shown below by Eqs.~\eqref{NCC2}--\eqref{order2}, no perturbative stability analysis is required to establish the emergence of singularities in these regular configurations. }

{\section{Interior evolution.}}
\noindent
The existence of regular BH solutions in Eq.~\eqref{finfi} presents two fundamental and intimately related challenges for their physical realization:
\begin{itemize}

\item Formation Mechanism: What processes could produce these non-singular spacetime geometries?

\item Evolutionary Fate: Once formed, does their evolution preserve regularity, or could singularities eventually develop?

\end{itemize}
To address these questions, while not claiming definitive resolution, we now investigate the {time dependent} extension of these solutions.
We begin by rewriting the metric~\eqref{metric} in Eddington-Finkelstein form by introducing the ingoing null coordinate $v$, defined through $dv=dt+dr/f$. We then promote the interior mass function, defined in the region $0< r \leq h$ [see Eq.~\eqref{mtransform}], to the fully general time-dependent form
\begin{equation}
	\label{promotting}
	m(r)\rightarrow\,m(v,r)\ ,
\end{equation}
which leads to the metric
\begin{equation}
	\label{EF-metric}
	ds^{2}
	=
	-f(v,r)\,dv^{2}+2\,dv\,dr+r^2\,d\Omega^2
\end{equation}
where
\begin{equation}
	\label{mtransform2}
{f(v,r)}=\left\{
	\begin{array}{l}
		1-\frac{2\,m(v,r)}{r}
		\ ,
		\quad
		{\rm for}\
		0< r \leq h
		\\
		\\
		1-\frac{2\,{\cal M}}{r}
		\ ,
		\quad
		{\rm for}\
			r\geq{h}
		\ .
	\end{array}
	\right.
\end{equation}
We emphasize that the total mass ${\cal M}$ remains entirely contained within the region $0<r\le h$. Consequently, the exterior geometry remains exactly Schwarzschild, implying the boundary condition
\begin{equation}
	\label{exterior}
	h=2{\cal M}\neq\,h(v)\ ,
\end{equation}
despite the ongoing evolution of the interior.

In the static case, the hypersurface $r=h$ is simultaneously an event horizon and a Killing horizon. The time-dependent extension in Eq.~\eqref{mtransform2}, however, changes the nature of this hypersurface and therefore of the entire spacetime. To characterize it, we employ the definition of a future outer trapping horizon (FOTH)~\cite{Hayward:1993wb,Ashtekar:2003hk},
\begin{equation}
	\label{FOTH}
	\theta_{(\ell)}=0\ ,\quad\theta_{(n)}<0\ ,\quad{\cal L}_n\theta_{(\ell)}<0\ ,
\end{equation} 
where the future-directed null vectors $\{\ell,n\}$ are 
\begin{equation}
	\ell^\mu=\frac{\partial}{\partial{v}}+\frac{f(v,r)}{2}\frac{\partial}{\partial{r}},\quad n^\mu=-\frac{\partial}{\partial{r}}\ .
\end{equation}
The corresponding null expansions are
\begin{equation}
	\theta_{(\ell)}=\frac{f(v,r)}{r}\ ,\quad\theta_{(n)}=-\frac{2}{r}<0\ ,
\end{equation} 
showing that every root of $f(v,r)=0$ defines a future marginally trapped surface. From Eq.~\eqref{mtransform2} it follows immediately that $r=h=2{\cal M}$ satisfies this condition. Moreover,
\begin{equation}
{\cal L}_n\theta_{(\ell)}\Big\rvert_{r=h}=-\frac{1}{h^2}<0\ ,
\end{equation} 
so that all conditions in Eq.~\eqref{FOTH} are fulfilled. Therefore, the hypersurface $r=h$ is a future outer trapping horizon.

The null convergence condition,
\begin{equation}
	R_{\mu\nu}\ell^\mu\ell^\nu\ge0\ ,
\end{equation}
then yields
\begin{equation}
	\label{NCC2}
	\dot m\ge0,
\end{equation}
where $\dot m\equiv\partial m/\partial v$. Because of the matching condition in Eq~\eqref{mtransform2} [see also~\eqref{exterior}], this inequality is saturated at the horizon,
\begin{equation}
	\dot m\Big|_{r=h}
	=
	\dot{\cal M}
	=
	0\ ,
\end{equation}
implying that no energy flux crosses $r=h$. Since the BH remains in equilibrium with its Schwarzschild exterior, the spacetime described by Eq.~\eqref{mtransform2} represents a BH with a dynamical interior enclosed by an isolated horizon, i.e., a non-expanding horizon in equilibrium~\cite{Ashtekar:1999yj,Ashtekar:2001jb}.

Finally, the formation of a Schwarzschild BH may be viewed schematically as the sequence
	\[
	\begin{array}{c}
		\text{(I) Gravitational collapse} \\
		\downarrow \\
		\text{(II) Dynamical horizon }  \\
		\text{(area increases)} \\
		\downarrow \\
		\text{(III) Isolated horizon } h = 2{\cal M} \\
		\text{(area fixed, Schwarzschild exterior)} \\
		\downarrow \\
		\text{(IV) Interior evolution } h_c = h_c(v) \\
		\downarrow \\
		\text{(V) Schwarzschild BH}
	\end{array}
	\]
	Our solution should therefore be regarded as an effective description of the BH interior after the horizon has formed, namely stages (III)-(V), rather than as a complete model of gravitational collapse. In this sense, it provides an analytic framework for the final stages of collapse, including the Schwarzschild limit obtained by setting $\beta_i=-1$ for some $i$ in Eq.~\eqref{n2(t)}.

Having established the geometric framework and its physical interpretation, we now investigate the time-dependent extension of the regular Schwarzschild solutions, with particular emphasis on the evolution and eventual decay of their de Sitter core.

\subsection{Quasi extremal $\leftrightarrow$ regular black holes}
\noindent 
{To analyze the evolution} of the nonsingular Schwarzschild geometries in Eq.~\eqref{finfi}, we first extend these solutions to time-dependent cases. This generalization requires promoting the mass function in Eq.~\eqref{minfi} via the generic prescription given in Eq.~\eqref{promotting}.

The constraint~\eqref{exterior} dictates that the {interior evolution} must be mediated exclusively through variations of the parameters $n_i$ in the mass function~\eqref{minfi}, yielding
\begin{equation}
	\label{ni(t)}
	n_i=n_i(v)\ ,
\end{equation}
which transforms the mass function to its time-dependent form
\begin{eqnarray}
	\label{minfi-v}
	&&	m(v,r)=\frac{r}{2}\left[\left(\frac{r}{h}\right)^2\,\prod_{i=1}^{N}\frac{n_i(v)+1}{n_i(v)-2}+3(-1)^N\times\right.\nonumber\\
	&&\left.\,\sum_{k=1}^{N}\frac{1}{n_k(v)-2}\left(\frac{r}{h}\right)^{n_k(v)}\prod_{\substack{i=1\\i\neq k}}^{N}\frac{n_i(v)+1}{n_k(v)-n_i(v)}\right]\ .
\end{eqnarray} 
Equation~\eqref{minfi-v} immediately reveals that the time dependent extension of Eq.~\eqref{minfi} develops singularities whenever $n_k(v)=n_i(v)$ for any $i\neq{k}$. To prevent these singularities, all $n_i(v)$ must evolve without intersections. This requires an initial ordering at $v=v_0$, i.e.,
\begin{equation}
	\label{order}
n_1(v_0)<n_2(v_0)<. . . <n_N(v_0)\ ,
\end{equation}
which must be preserved for all $v$ through the constraint
\begin{equation}
	\label{order2}
	\dot{n}_i(v)\leq\dot{n}_j(v)\ ,\quad\forall\,\, i<j\ .
\end{equation}
[The reversed ordering similarly requires $\dot{n}_i(v)\geq\dot{n}_j(v)$.] Eqs.~\eqref{order} and~\eqref{order2} demonstrate that the evolution of the Schwarzschild BH interior is highly constrained: only synchronized parameter evolution yields physically viable configurations {{ [see Fig.~\ref{fig4}]}. On the other hand, the constraint in Eq.~\eqref{order2} contains both negative and positive slopes ($\dot{n}_i < 0$ and $\dot{n}_i > 0$), which correspond physically to distinct evolutionary regimes: negative slopes drive collapse while positive slopes lead to expansion. When collapse ($\dot{n}_i < 0$)  dominates, the system evolves from an extremal configuration to a regular BH; conversely, expansion ($\dot{n}_i > 0$) dominance produces the reverse transition from regular to extremal configurations [as described for the static case by Eq.~\eqref{extremal}]. This formulation reveals that the inner horizon $h_{\rm c}$ becomes a {time dependent} hypersurface $h_{\rm c} = h_{\rm c}(v)$, as illustrated in Fig.~\ref{fig1}. 

\begin{figure}
	\includegraphics[width=0.45\textwidth]{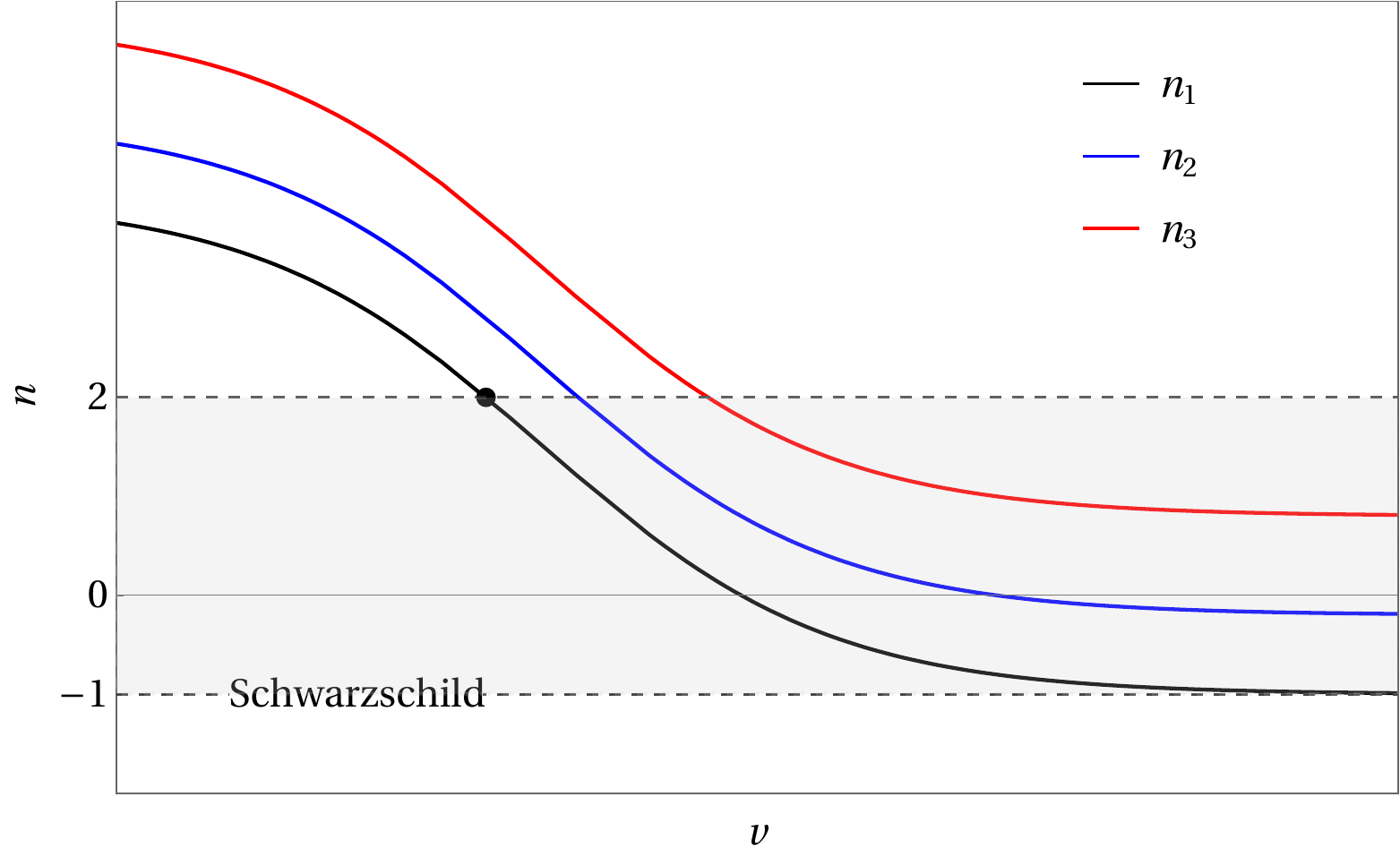}  \
		\includegraphics[width=0.45\textwidth]{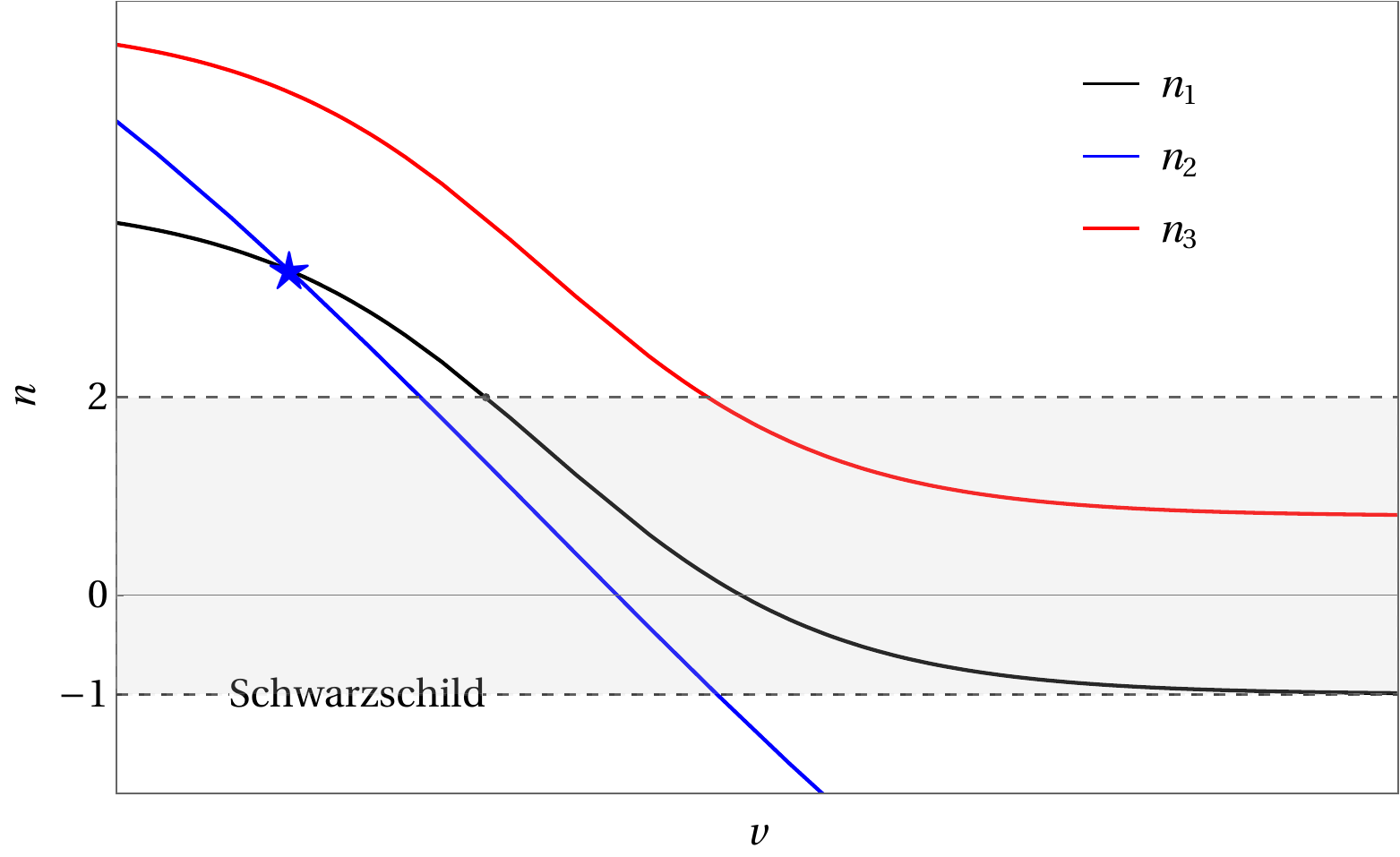}  \
\caption{\footnotesize {Case $N=3$. Schematic evolution of the three functions $\{n_1,n_2,n_3\}$ in Eq.~\eqref{minfi-v}. Top: The functions satisfy the constraints in Eqs.~\eqref{order} and~\eqref{order2}. The parameter $n_1$ is the first to reach the critical threshold $n_1=2$ (black point), and eventually attains the Schwarzschild value $n_1=-1$. Bottom: Since $\dot n_1>\dot n_2$, the condition in Eq.~\eqref{order2} is violated, causing the trajectories of $n_1$ and $n_2$ to intersect (blue star), where the solution becomes singular before the threshold $n_i=2$ is reached. This illustrates that Eq.~\eqref{order2} provides a simple sufficient condition for preserving the ordering. The shaded region denotes the singular interval $n_i\in[-1,2]$.}}
	\label{fig4}
\end{figure}

\begin{figure}
	\includegraphics[width=0.45\textwidth]{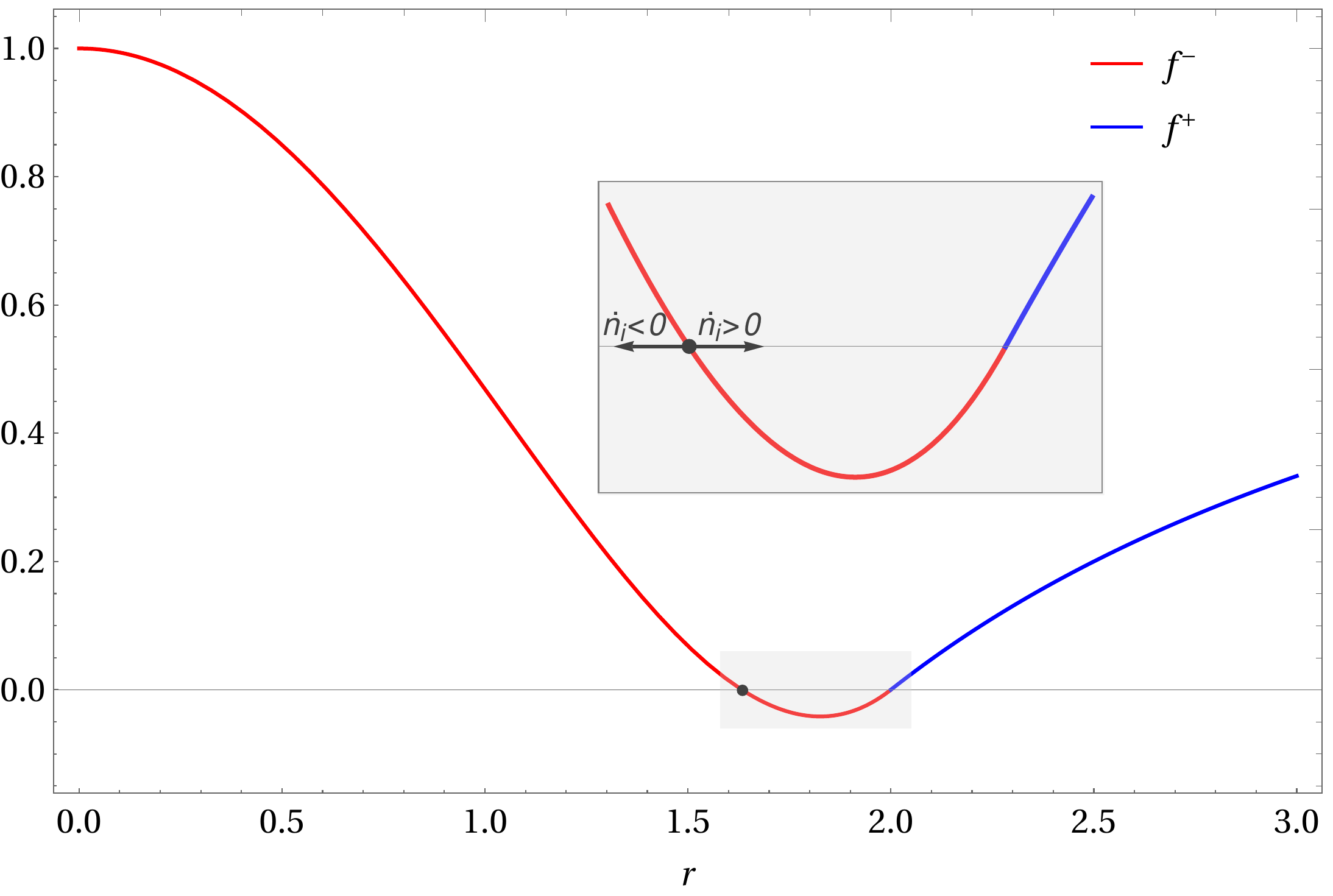}  \
	\caption{\footnotesize Evolution of the inner horizon $h_{\rm c}(v)$ for the {interior BH geometry defined by the time dependent mass function~\eqref{minfi-v}}: Collapse for $\dot{n}_i<0$ and expansion for $\dot{n}_i>0$. Horizon $h=2{\cal M}\neq\,h(v)$. Distance $r$ in units of ${\cal M}$. }
	\label{fig1}
\end{figure}
Modeling the formation of an {quasi} extremal configuration emerging from a regular BH is certainly appealing. However, this scenario carries a significant physical limitation: it entails violating the null convergence condition in~\eqref{NCC2}. By contrast, in collapse-dominated regimes ($\dot{n}_i < 0$), Eq.~\eqref{NCC2} is trivially satisfied, indicating that the only physically viable {classical} pathway is a transition from a {quasi extremal BH}.  Moreover, the geometric condition $G_{\alpha\beta}u^{\alpha}u^{\beta}\geq\,0$ for all future-directed timelike $u^{\alpha}$, which corresponds to the weak energy condition within general relativity, always holds. 
This persistent collapse inertia drives the formation of a secondary horizon, resulting in a regular BH configuration whose evolution continues until it reaches the critical threshold $n_i(v)=2$ in Eq.~\eqref{minfi-v}, the precise condition for singularity formation, where the Kretschmann scalar behaves as $\sim[\text{Log}(r/h)]^2$ for $r\sim0$. 

{What is the causal character of this singularity? For the line element~\eqref{EF-metric}, the normal to a hypersurface of constant $r$ satisfies
	\begin{equation}
		n_\mu n^\mu=g^{rr}=f(v,r).
	\end{equation}
	Hence, a singularity at $r=0$ is timelike when $f(v,0^+)>0$, since its normal is spacelike, and spacelike when $f(v,0^+)<0$, since its normal is timelike. As shown in Fig.~\ref{fig1}, collapse leads to the shrinking of the inner horizon $h_{\rm c}(v)$, which eventually disappears. As long as $h_{\rm c}(v)>0$, the region near $r=0$ satisfies $f(v,0^+)>0$, and the singularity is therefore timelike. When $h_{\rm c}(v)$ disappears, one finds $f(v,0^+)<0$, and the singularity becomes spacelike. This causal transition is thus directly associated with the disappearance of the inner horizon. See Table~\ref{tab} for details and the companion paper~\cite{Ovalle:2026lxb}, which contains the detailed analysis of Schwarzschild singularity formation. In particular, the transition between these two regimes involves the loss of a locally Minkowskian neighborhood at $r=0$, a phenomenon that we have termed ``Minkowski breaking''~\cite{Ovalle:2026lxb}.}

We conclude this section by emphasizing that the metric~\eqref{EF-metric} together with~\eqref{promotting} is well established both within general relativity~\cite{Husain:1995bf,Wang:1998qx} and in more general gravitational frameworks~\cite{Borissova:2025msp}. Within this setting, the mass function~\eqref{minfi-v} generates a configuration characterized by a nonuniform energy density $\rho=\rho(v,r)$, intrinsic anisotropy $\Delta\equiv p_\theta-p_r\neq 0$, nonvanishing radial pressure $p_r(v,r)$ and tangential pressure $p_\theta(v,r)$, as well as a nonzero energy flux $\epsilon=\epsilon(v,r)$. From this standpoint, and in direct comparison with earlier analyses, it becomes clear that the nonuniform and anisotropic collapsing fluid described here lies far outside the highly constrained matter models traditionally used to investigate BH interiors and singularity formation. In particular, it cannot be captured by homogeneous dust collapse~\cite{Oppenheimer:1939ue,Shibata:1999va}, inhomogeneous dust collapse~\cite{Tolman:1934za,Bondi:1947fta,Lemaitre:1933gd,Christodoulou:1984mz,Joshi:1993zg,Joshi:2001xi}, nor by isotropic~\cite{Lasky:2006mg,Mosani:2020ena,Joshi:2023ugm} or anisotropic configurations~\cite{Mena:2004ck} that do not belong to the Kerr–Schild class. This establishes our framework as a genuinely broader and more general setting for probing the interior evolution of BHs and the onset of singular behavior. {In particular, as we will see in the next section, our framework allows a bounce to be introduced quite naturally as a regularization mechanism [see Eq.~\eqref{bounce} below], as occurs in several quantum-gravity-inspired scenarios~\cite{Malafarina:2017csn,Hergott:2022hjm}. Such a bounce requires a violation of the null convergence condition, suggesting that the generic fluid $\{\rho,p_r,p_{\theta},\epsilon\}$ may be interpreted as an effective description of corrections beyond classical matter. In this sense, the present construction does not rely on a specific microscopic matter model, but provides a geometric framework in which such corrections can be effectively encoded.}
\section{Singularity resolution\\and final remarks.}
\label{sec3}
\noindent
Our analysis reveals that singularities generically emerge in regular Schwarzschild BHs bounded by an isolated horizon, as an exact consequence of Eqs.~\eqref{NCC2}--\eqref{order2}, without assuming any particular gravitational theory or resorting to perturbative methods. Avoiding this outcome would require extending these equations with additional, ad hoc constraints. Nevertheless, such modifications may offer valuable insights into mechanisms for singularity resolution. As a concrete example, we propose a saturating transition of the form
\begin{equation}
	\label{n2(t)}
n_i(v)=\frac{\alpha_i+\beta_i}{2}+\frac{\beta_i-\alpha_i}{2}\tanh(\omega{v})\ , \quad \forall\,i\ , 
\end{equation}
where $\alpha_i\equiv\,n_i(-\infty)$ and $\beta_i\equiv\,n_i(+\infty)$ characterize the initial and final BH states, respectively, and $\tau\equiv\omega^{-1}$ sets the transition timescale. The specific transition profile is not essential: any smooth monotonic interpolation between the prescribed asymptotic states $\{\alpha_i,\beta_i\}$ yields the same qualitative evolution, provided the ordering constraints in Eqs.~\eqref{order} and~\eqref{order2} are satisfied. The quasi extremal to regular transition demands
\begin{equation}
	\label{conditions}
	1\ll\alpha_i\gg\beta_i>2\ ,\quad\forall\,\, i\ .
\end{equation} 
We stress that $\alpha_i$ are unrestricted: choosing $\alpha_i \gg 1$ allows arbitrarily close approach to the initial extremal configuration
\begin{equation}
	\label{extremal2}
	m(-\infty,r) \sim \frac{r^3}{2h^2}, \quad
	f(-\infty,r) \sim 1 - \frac{r^2}{h^2} 
\end{equation}
while still preserving horizon differentiability, as illustrated in Fig.~\ref{fig3}. This model fully captures the transition from quasi-extremal to regular configurations, providing a detailed description of regular BH formation with high resolution. Conversely, the reverse transition, from regular to quasi-extremal, necessarily entails a violation of the null convergence condition~\eqref{NCC2}, and can be obtained simply by interchanging $\alpha\leftrightarrow\beta$ in Eq.~\eqref{conditions}. Moreover, the model also reproduces the formation of Schwarzschild's singularity with remarkable clarity, which follow directly by setting $\beta_i=-1$ for some $i$ in Eq.~\eqref{conditions}. Finally, one may introduce a  second resolution: avoiding the threshold $n_i(v)=2$ through a bounce during collapse~\cite{BenAchour:2020gon}, a process that again requires violation of the null convergence condition. This scenario can be modeled by 
\begin{equation}
	\label{bounce}
	n_i(v)=\alpha_i-\beta_i\,e^{-(\omega\,v)^2}\ ,
\end{equation}
\begin{figure}
	\includegraphics[width=0.45\textwidth]{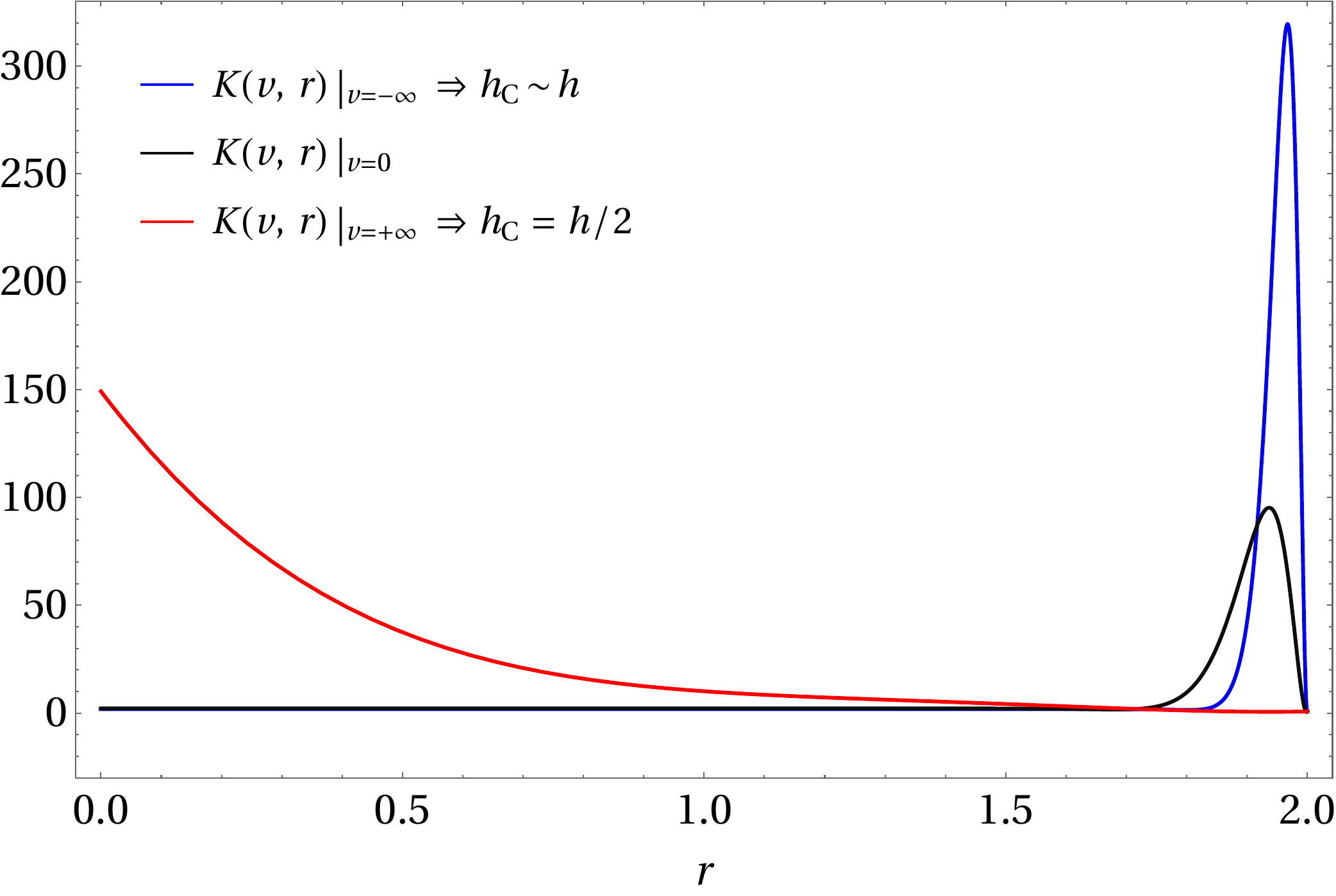}  \
	\caption{\footnotesize Kretschmann $K(v,r)$ from a quasi extremal to a regular BH for $N=2$ in~\eqref{Mr}, where $n(v)$ is given by~\eqref{n2(t)} with $\alpha=62$, $\beta=3$ and $l(v)=1+n(v)$. There is a maximum just below the horizon: $K(-\infty,h-\delta)\sim(\alpha^2/h^5)\delta$, with $\delta\ll\,1.$ We take $\omega=0.1$.  }
	\label{fig3}
\end{figure}
with $\alpha_i\gtrsim\beta_i\gg\,1$. While the first model in Eq.~\eqref{n2(t)} circumvents the need for weak cosmic censorship, the second model in~\eqref{bounce} instead reinterprets the event horizon as a boundary that conceals a transient violation of the null convergence condition, a process essential for realizing a non-singular configuration. {Here we emphasize that violation of the null convergence condition does not invalidate these models per se. Rather, it indicates that the corresponding processes cannot occur while preserving the classical convergence condition. Violations of classical energy conditions are known to arise in semiclassical and quantum settings, and may therefore occur, at least locally, in regimes where quantum effects become relevant~\cite{Visser:1995cc,Fewster:2012yh,Kontou:2020bta}.}

Finally, we stress, following Ref.~\cite{Borissova:2025msp}, that evolving inner horizons need not always coincide with Cauchy horizons. In contrast, our model shows that the inner horizon $h_c(v)$, although time dependent, remains a genuine Cauchy horizon as long as the configuration is nonsingular, that is, for $n(v)>2$. This conclusion follows directly from the Penrose singularity theorem: since a trapped surface exists at $r=h$ and the null convergence condition holds throughout the kinematical process, the single inner horizon $h_c(v)$ must be a Cauchy horizon to evade the theorem. If a regularization mechanism such as that in Eq.~\eqref{n2(t)} is imposed, the inner horizon is expected to become unstable under generic perturbations~\cite{Penrose:1968,Poisson:1989zz,Ori:1991zz,Carballo-Rubio:2018pmi,Carballo-Rubio:2021bpr,Carballo-Rubio:2024dca}. Nevertheless, we emphasize that the emergence of singularities is established without any perturbative analysis and follows directly from the exact time-dependent solutions presented here. Our conclusions are therefore rigorously valid within the class of geometries considered, namely the Kerr-Schild subclass ($g_{tt}g_{rr}=-1$) with an isolated horizon ($\dot{\cal M}=0$). Whether similar results persist in more general BH spacetimes, particularly beyond spherical symmetry, remains an important open problem. We wish to conclude by emphasizing that the existence of a regular BH as the final outcome of collapse is only possible if the collapsing configuration successfully avoids singularities throughout its formation. This requirement inevitably constrains the process and, as demonstrated by Eqs.~\eqref{order} and~\eqref{order2}, may be far more restrictive than anticipated. Exploring how such constraints manifest in more realistic scenarios, such as rotational collapse, could not only clarify the precise conditions for singularity resolution but also shed light on the fundamental interplay between geometry, energy conditions, and cosmic censorship.

\begin{center}
	\textit{``I fumi non fanno ombre terminate, ed i loro confini sono tanto meno noti, quanto essi sono più distanti dalle loro cause\ldots''}
	
	\textit{--- Leonardo da Vinci, a proposito dello sfumato}
\end{center}

\subsection*{Acknowledgments}
\vspace*{1mm}
JO thanks YITP for all their support during the development of this research work. This work is partially supported by ANID
FONDECYT Grant No. 1250227.
%

%
%
%
\bibliography{references}
\bibliographystyle{apsrev4-1}
%
%
\end{document}